\documentclass[a4paper,11pt]{article}
\pdfoutput=1 

\usepackage{jheppub} 

\usepackage[T1]{fontenc} 

\title{Higgs inflation and scalar weak gravity conjecture}

\author{Yang Liu}

\affiliation[a]{School of Physics and Astronomy, University of Nottingham, Nottingham NG7 2RD, UK}
\affiliation[b]{Nottingham Centre of Gravity, University of Nottingham, Nottingham NG7 2RD, UK}

\emailAdd{yang.liu@nottingham.ac.uk}

\abstract{In this article, we intend to find a specific model which can satisfy the further refining dS swampland conjecture and scalar weak gravity conjecture (SWGC) simultaneously, in particular, Higgs inflation model and its two extensions: Higgs-Dilaton model and Palatini Higgs inflation. We find that although Higgs inflation model and Higgs-Dilaton model could satisfy the further refining dS swampland conjecture, the two models cannot satisfy SWGC and its strong version. While Palatini Higgs inflation could satisfy these conjectures simultaneously. Therefore Palatini Higgs inflation could be a “real” inflation model of the universe. }

\begin{document} 
\maketitle
\flushbottom

\section{Introduction}
\label{sec:intro}
The swampland program is a very interesting development for the phenomenology of quantum gravity theories. In recent years, people have proposed the “de Sitter swampland conjecture” and “refined de Sitter swampland conjecture” considering the derivative of scalar field potentials, such as [1,2]. If we consider a $4d$ theory of real scalar field $\phi^i$ coupled to gravity, whose dynamics is governed by a scalar potential $V(\phi^j)$, then the action is given by [2,3]
\begin{equation}\label{eq:1.1}
S = \int_{4} \sqrt{|g_4|} \left(\frac{M^2_p}{2} R_4 - \frac{1}{2} h_{ij} \partial_{\mu} \phi^i \partial^{\mu} \phi^j - V \right),
\end{equation}
where $M_p$ is the Planck mass, $|g_4|$ is the determinant of the metric matrix of $4d$ spacetime, $R_4$ is the $4d$ Riemann curvature of spacetime and $h_{ij}$ is the metric on the target space of the scalar fields. \\
Firstly, we review the refined de Sitter conjecture. The conjecture states that an effective theory of quantum gravity, which is not in the swampland, should at least meet one of the following two constraints [1,2]:
\begin{equation}\label{eq:1.2}
|\nabla V| \geq \frac{c_1}{M_p} \cdot V,
\end{equation}
or
\begin{equation}\label{eq:1.3}
min(\nabla_i \nabla_j V) \leq -\frac{c_2}{M^2_p} \cdot V,
\end{equation}
where $c_1$ and $c_2$ are both positive constants of the order of 1. The first constraint, i.e., eq.$(1.2)$, corresponds to original “swampland conjecture”.\\
Furthermore, David Andriot and Christoph Roupec proposed a further refining de Sitter swampland conjecture, which suggested that a low energy effective theory of a quantum gravity that takes the form $(1.1)$ should satisfy [2,3], 
\begin{equation}\label{eq:1.4}
\left(M_p \frac{|\nabla V|}{V} \right)^q - a M^2_p \frac{min(\nabla_i \nabla_j V)}{V} \geq b \qquad with \quad a+b=1, \quad a,b>0 \quad q>2.
\end{equation}
In our previous work, we have found that Higgs inflation model and its two extensions: Higgs-Dilaton model and Palatini Higgs inflation all satisfy the refining de Sitter swampland conjecture [3]. \\
On the other hand, Palti proposed a generalisation of the Weak Gravity Conjecture in the presence of scalar fields (SWGC) [4]. Moreover, Strong Scalar Weak Gravity Conjecture (SSWGC) has been proposed by Eduardo Gonzalo and Luis E.Ibáñez [5]. The potential of scalar field $V(\Phi)$, which is coupled to quantum gravity, should meet certain conditions. We will review SWGC and SSWGC briefly in section 3. The main goal of this work is to find a specific inflation model which satisfy the two versions of SWGC and the further refining de Sitter swampland conjecture simultaneously, in particular, Higgs inflation model and its two extensions: Higgs-Dilaton model and Palatini Higgs inflation. \\
The article is composed as follows: in section 2, we review Higgs inflation model and its two extensions briefly. In section 3, we review the two versions of scalar weak gravity conjecture and examine if these inflation models can satisfy the conjectures. In section 4, the results we have obtained are discussed.

\section{Higgs inflation and its extensions}
In this section, we will review the Higgs inflation model and its two extensions briefly: Higgs-Dilaton model and Palatini Higgs inflation [6].

\subsection{Higgs inflation model}
The total action of Higgs inflation is 
\begin{equation}\label{eq:2.1}
S = \int d^4 x \sqrt{-g} [\frac{M^2_p}{2}R + \xi H^{\dagger} H R + L_{SM}],
\end{equation}
which contains two parameters, namely, the Higgs mass expectation value $v_{EW}\approx 250GeV$ and the reduced Planck mass $M_p = 2.435 \times 10^{18}GeV$ [6].  When the value of field is very large, the Planck mass plays an important role for inflation [6,7,8]. If we take the unitary gauge, Higgs field can be written as $H= (0, h)^T/\sqrt{2}$ [6]. Then the action eq.$(2.1)$ can be rewritten as 
\begin{equation}\label{eq:2.2}
S = \int d^4 x \sqrt{-g} [\frac{M^2_p + \xi h^2}{2}R - \frac{1}{2} (\partial h)^2 - U(h)],
\end{equation}
where
\begin{equation}\label{eq:2.3}
U(h) = \frac{\lambda}{4} (h^2 - v^2_{EW})^2
\end{equation} 
is the symmetry breaking potential in the Standard Model [6,7,8].\\
The action $(2.2)$ can be reformulated in the Einstein frame by a Weyl transformation $g_{\mu\nu} \rightarrow \Theta g_{\mu\nu}$ with [6]:
\begin{equation}\label{eq:2.4}
\Theta^{-1} = 1 + \frac{h^2}{F^2_{\infty}}, \qquad F_{\infty} \equiv \frac{M_p}{\sqrt{\xi}}.
\end{equation}
Then action $(2.2)$ in the Einstein frame, can be rewritten as [3,6,7,8]:
\begin{equation}\label{eq:2.5}
S = \int d^4 x \sqrt{-g} [\frac{M^2_p}{2}R - \frac{1}{2} M^2_p K(\Theta) (\partial \Theta)^2 - V(\Theta)],
\end{equation}
which contains a non-canonical kinetic sector [3,6]:
\begin{equation}\label{eq:2.6}
K(\Theta) \equiv \frac{1}{4 |a| \Theta^2} \left(\frac{1-6|a|\Theta}{1-\Theta} \right), 
\end{equation}
where
\begin{equation}\label{eq:2.7}
a \equiv -\frac{\xi}{1+6\xi}, 
\end{equation}
and a non-exactly flat potential [3,6]:
\begin{equation}\label{eq:2.8}
V(\Theta) \equiv U(\Theta) \Theta^2 = \frac{\lambda F^4_{\infty}}{4} [1- \left(1+ \frac{v^2_{EW}}{F^2_{\infty}} \right) \Theta]^2.
\end{equation}

\subsection{Higgs-Dilaton model}
The existence of robust predictions in (non-critical) Higgs inflation is intimately related to the emerging dilatation symmetry of its tree-level action at large field values [3]. The uplifting of Higgs inflation to a completely scale-invariant setting was considered in several articles [9,10,11,12,13]. If we take the unitary gauge $H= (0, h)^T/\sqrt{2}$, the action of the graviscalar sector of the Higgs-dilaton model is given by
\begin{equation}\label{eq:2.9}
S = \int d^4 x \sqrt{-g} [\frac{\xi_h h^2 + \xi_{\chi} \chi^2}{2}R - \frac{1}{2} (\partial h)^2 - \frac{1}{2} (\partial \chi)^2 - V(h, \chi)], 
\end{equation}
where
\begin{equation}\label{eq:2.10}
U(h,\chi)= \frac{\lambda}{4} (h^2 - \alpha \chi^2)^2 + \beta \chi^4 
\end{equation}
is a scale-invariant version of the Standard Model symmetry breaking potential and $\alpha$, $\beta$ positive dimensionless parameters [3,6,9,10,11,12,13].\\
The action in the Einstein frame can be obtained by a Weyl rescaling $g_{\mu\nu} \rightarrow M^2_p/(\xi_h h^2 + \xi_{\chi} \chi^2)g_{\mu\nu}$ and a field redefinition, namely,
\begin{equation}\label{eq:2.11}
S = \int d^4 x \sqrt{-g} [\frac{M^2_p}{2}R - \frac{1}{2} M^2_p K(\Theta) (\partial \Theta)^2 - \frac{1}{2} \Theta (\partial \Phi)^2 - U( \Theta)]. 
\end{equation}
It contains a kinetic sector for $\Theta$ field
\begin{equation}\label{eq:2.12}
K(\Theta) = \frac{1}{4 |\bar{a}| \Theta^2} \left(\frac{c}{|\bar{a}|\Theta - c} + \frac{1-6|\bar{a}|\Theta}{1-\Theta} \right),
\end{equation}
which has two “inflationary” poles at $\Theta =0$ and $\Theta = c/|\bar{a}|$ and a “Minkowski” pole at $\Theta =1$ [6] (The “Minkowski” pole does not play a significant role during inflation and can be neglected for all practical purposes [6]) and a potential
\begin{equation}\label{eq:2.13}
U(\Theta) = U_0 (1-\Theta)^2, \qquad U_0 \equiv \frac{\lambda M^4_p}{4} \left(\frac{1 + 6\bar{a}}{\bar{a}} \right)^2, 
\end{equation}
where
\begin{equation}\label{eq:2.14}
a \equiv - \frac{\xi_h}{1+6\xi_h}, \qquad \bar{a} \equiv a \left(1- \frac{\xi_h}{\xi_{\chi}} \right). 
\end{equation}

\subsection{Palatini Higgs model}
The action is minimized with respect to the metric in Higgs inflation model. This procedure implicitly assumes the inclusion of a York-Hawking-Gibbons term ensuring the cancellation of a total derivative term with no-vanishing variation at the boundary and the existence of a Levi-Civita connection depending on the metric tensor [6,14,15]. Then one could consider a Palatini formulation of gravity alternatively. In this formulation, no additional boundary term is required and the metric tensor and the connection are treated independently [6,14,15]. We consider the action $(2.2)$ with $R=g^{\mu\nu}R_{\mu\nu}(\Gamma, \partial\Gamma)$ and $\Gamma$ a non-Levi-Civita connection in order to see explicitly. At $\phi \gg v_{EW}$, the action in the Einstein frame can be obtained by a Weyl transformation $g_{\mu\nu} \rightarrow \Theta g_{\mu\nu}$ and a field redefenition [3,6],
\begin{equation}\label{eq:2.15}
S = \int d^4 x \sqrt{-g} [\frac{M^2_p}{2}R - \frac{1}{2} (\partial \phi)^2 - V(\phi)], 
\end{equation}
with
\begin{equation}\label{eq:2.16}
V(\phi) = \frac{\lambda}{4} F^4(\phi), \qquad F(\phi) \equiv F_{\infty} \tanh \left(\frac{\sqrt{a}\phi}{M_p} \right) 
\end{equation}
One can find more details of this model in ref.[6].

\section{Higgs inflation and scalar weak gravity conjecture}
In this section, we will check if Higgs inflation model and its two extensions satisfy scalar weak gravity conjecture. Firstly, we will review scalar weak gravity conjecture and its strong version briefly.\\
Palti formulated a first version of a Scalar Weak Gravity Conjecture (SWGC) [4]. Considering a particle H with mass $m$ which is coupled to a light scalar $\phi$ with a triliniear coupling proportional to $\mu = \partial_{\phi} m$. SWGC states that the force mediated by $\phi$ must be stronger than gravitational force and $m^2(\phi)$ is considered as a function of $\phi$ so that $m^2= V''$, then we have
\begin{equation}\label{eq:3.1}
(V^{(3)})^2 \geq \frac{(V^{(2)})^2}{M^2_p}.
\end{equation} 
Furthermore, Eduardo Gonzalo and Luis E.Ibáñez proposed a strong version of SWGC, i.e., SSWGC [5]. The conjecture states that the potential of any canonically normalized real scalar $V(\phi)$ in the theory must satisfy for any value of the field the constraint:
\begin{equation}\label{eq:3.2}
2(V^{(3)})^2 - V^{(2)} V^{(4)} \geq \frac{(V^{(2)})^2}{M^2_p},
\end{equation} 
with primes denoting derivation with respect to $\phi$.\\
Then, we will check if Higgs inflation model and its two extensions satisfy scalar weak gravity conjecture. In the following section, we take $M_p=1$.

\subsection{Higgs inflation model}
From eq.$(2.8)$, we can obtain
\begin{equation}\label{eq:3.3}
V^{(1)}(\Theta) = \frac{-\lambda F^4_{\infty}}{2} (1+ \frac{v^2_{EW}}{F^2_{\infty}}) [1- (1+\frac{v^2_{EW}}{F^2_{\infty}}) \Theta], 
\end{equation}
\begin{equation}\label{eq:3.4}
V^{(2)}(\Theta) = \frac{\lambda F^4_{\infty}}{2} (1+ \frac{v^2_{EW}}{F^2_{\infty}})^2, 
\end{equation}
\begin{equation}\label{eq:3.5}
V^{(3)}(\Theta) = V^{(4)}(\Theta) = 0.
\end{equation}
Then we have,
\begin{equation}\label{eq:3.6}
(V^{(2)})^2 > 0,
\end{equation}
while
\begin{equation}\label{eq:3.7}
(V^{(3)})^2=0,
\end{equation}
\begin{equation}\label{eq:3.8}
2 (V^{(3)})^2 - V^{(2)}  V^{(4)} = 0.  
\end{equation}
Therefore, Higgs inflation model satisfies neither scalar weak gravity conjecture nor strong scalr weak gravity conjecture.

\subsection{Higgs-Dilaton model}
From eq.(2.13),  we can obtain
\begin{equation}\label{eq:3.9}
U^{(1)}(\Theta) = 2 U_0 (\Theta - 1), 
\end{equation}
\begin{equation}\label{eq:3.10}
U^{(2)}(\Theta) = 2 U_0, 
\end{equation}
\begin{equation}\label{eq:3.11}
U^{(3)}(\Theta) = U^{(4)}(\Theta) = 0.
\end{equation}
Then we have,
\begin{equation}\label{eq:3.12}
(U^{(2)})^2 > 0,
\end{equation}
while
\begin{equation}\label{eq:3.13}
(U^{(3)})^2=0,
\end{equation}
\begin{equation}\label{eq:3.14}
2 (U^{(3)})^2 - U^{(2)}  U^{(4)} = 0.  
\end{equation}
Therefore, Higgs-Dilaton model satisfies neither scalar weak gravity conjecture nor strong scalr weak gravity conjecture.

\subsection{Palatini Higgs model}
From eq.(2.16), we can obtain
\begin{equation}\label{eq:3.15}
V^{(1)}(\phi) = \lambda \sqrt{a} F^4_{\infty} \frac{\sinh^3 (\sqrt{a} \phi)}{\cosh^5 (\sqrt{a} \phi)},  
\end{equation}
\begin{equation}\label{eq:3.16}
V^{(2)}(\phi) = \lambda a F^4_{\infty} [\frac{3 \sinh^2 (\sqrt{a} \phi)}{\cosh^4 (\sqrt{a} \phi)} - \frac{5 \sinh^4 (\sqrt{a} \phi)}{\cosh^6 (\sqrt{a} \phi)} ] \equiv \frac{\lambda a F^4_{\infty}}{\cosh^6 (\sqrt{a} \phi)}  A,  
\end{equation}
where
\begin{equation}\label{eq:3.17}
A = 3 \sinh^2 (\sqrt{a} \phi)\cosh^2 (\sqrt{a} \phi) - 5 \sinh^4 (\sqrt{a} \phi).  
\end{equation}
Then
\begin{equation}\label{eq:3.18}
V^{(3)}(\phi)= -\frac{6\lambda a^{3/2} F^4_{\infty}}{\cosh^7 (\sqrt{a} \phi)} \sinh(\sqrt{a} \phi) A + \frac{\lambda a F^4_{\infty}}{\cosh^6 (\sqrt{a} \phi)} A'.  
\end{equation}
\begin{equation}\label{eq:3.19}
\begin{aligned}
V^{(4)}(\phi)= &\frac{42\lambda a^{2} F^4_{\infty}}{\cosh^8 (\sqrt{a} \phi)} \sinh^2(\sqrt{a} \phi) A - \frac{6\lambda a^2 F^4_{\infty}}{\cosh^6 (\sqrt{a} \phi)} A \\
& - \frac{12 \lambda a^{3/2} F^4_{\infty}}{\cosh^7 (\sqrt{a} \phi)} \sinh(\sqrt{a} \phi) A' + \frac{\lambda a F^4_{\infty}}{\cosh^6 (\sqrt{a} \phi)} A''.\\  
\end{aligned}
\end{equation}
Thus we have
\begin{equation}\label{eq:3.20}
(V^{(2)})^2= \frac{\lambda^2 a^2 F^8_{\infty}}{\cosh^{12} (\sqrt{a} \phi)} A^2, 
\end{equation}
\begin{equation}\label{eq:3.21}
(V^{(3)})^2= \frac{36 \lambda^2 a^3 F^8_{\infty}}{\cosh^{14} (\sqrt{a} \phi)} \sinh^2 (\sqrt{a} \phi) A^2 + \frac{\lambda^2 a^2 F^8_{\infty}}{\cosh^{12} (\sqrt{a} \phi)} A'^2 - \frac{12 \lambda^2 a^{5/2} F^8_{\infty}}{\cosh^{13} (\sqrt{a} \phi)} \sinh (\sqrt{a} \phi) A A', 
\end{equation}
\begin{equation}\label{eq:3.22}
\begin{aligned}
V^{(2)}+ V^{(4)}= &\frac{\lambda a F^4_{\infty}}{\cosh^6 (\sqrt{a} \phi)}  A + \frac{42\lambda a^{2} F^4_{\infty}}{\cosh^8 (\sqrt{a} \phi)} \sinh^2(\sqrt{a} \phi) A - \frac{6\lambda a^2 F^4_{\infty}}{\cosh^6 (\sqrt{a} \phi)} A \\
& - \frac{12 \lambda a^{3/2} F^4_{\infty}}{\cosh^7 (\sqrt{a} \phi)} \sinh(\sqrt{a} \phi) A' + \frac{\lambda a F^4_{\infty}}{\cosh^6 (\sqrt{a} \phi)} A''.\\  
\end{aligned}
\end{equation}
and
\begin{equation}\label{eq:3.23}
\begin{aligned}
V^{(2)} (V^{(2)}+ V^{(4)})= &\frac{\lambda^2 a^2 F^8_{\infty}}{\cosh^{12} (\sqrt{a} \phi)}  A^2 + \frac{42\lambda^2 a^{3} F^8_{\infty}}{\cosh^{14} (\sqrt{a} \phi)} \sinh^2(\sqrt{a} \phi) A^2 - \frac{6\lambda^2 a^3 F^8_{\infty}}{\cosh^{12} (\sqrt{a} \phi)} A^2 \\
& - \frac{12 \lambda^2 a^{5/2} F^8_{\infty}}{\cosh^{13} (\sqrt{a} \phi)} \sinh(\sqrt{a} \phi) A A' + \frac{\lambda^2 a^2 F^8_{\infty}}{\cosh^{12} (\sqrt{a} \phi)} A A''.\\  
\end{aligned}
\end{equation}
If Palatini Higgs inflation satisfies SWGC, then based on eq.(3.1), (3.20) and (3.21), we have
\begin{equation}\label{eq:3.24}
\frac{36 \lambda^2 a^3 F^8_{\infty}}{\cosh^{14} (\sqrt{a} \phi)} \sinh^2 (\sqrt{a} \phi) A^2 + \frac{\lambda^2 a^2 F^8_{\infty}}{\cosh^{12} (\sqrt{a} \phi)} A'^2 - \frac{12 \lambda^2 a^{5/2} F^8_{\infty}}{\cosh^{13} (\sqrt{a} \phi)} \sinh (\sqrt{a} \phi) A A' - \frac{\lambda^2 a^2 F^8_{\infty}}{\cosh^{12} (\sqrt{a} \phi)} A^2 \geq 0. 
\end{equation}
If Palatini Higgs inflation satisfies SSWGC, then based on eq.(3.2), (3.21) and (3.23), we have
\begin{equation}\label{eq:3.25}
\begin{aligned}
\frac{72 \lambda^2 a^3 F^8_{\infty}}{\cosh^{14} (\sqrt{a} \phi)} \sinh^2 (\sqrt{a} \phi) A^2 + \frac{ 2\lambda^2 a^2 F^8_{\infty}}{\cosh^{12} (\sqrt{a} \phi)} A'^2 - \frac{24 \lambda^2 a^{5/2} F^8_{\infty}}{\cosh^{13} (\sqrt{a} \phi)} \sinh (\sqrt{a} \phi) A A' \geq \\
\frac{\lambda^2 a^2 F^8_{\infty}}{\cosh^{12} (\sqrt{a} \phi)}  A^2 + \frac{42\lambda^2 a^{3} F^8_{\infty}}{\cosh^{14} (\sqrt{a} \phi)} \sinh^2(\sqrt{a} \phi) A^2 - \frac{6\lambda^2 a^3 F^8_{\infty}}{\cosh^{12} (\sqrt{a} \phi)} A^2 \\
-\frac{12 \lambda^2 a^{5/2} F^8_{\infty}}{\cosh^{13} (\sqrt{a} \phi)} \sinh(\sqrt{a} \phi) A A' + \frac{\lambda^2 a^2 F^8_{\infty}}{\cosh^{12} (\sqrt{a} \phi)} A A''.\\  
\end{aligned}
\end{equation}
Eq.(3.24) and (3.25) are the constraints of Palatini Higgs inflation which meet the two versions of scalar weak gravity conjecture.

\section{Conclusions and discussions}
In ref.[4], Palti proposed a first version of a Scalar Weak Gravity Conjecture (SWGC). SWGC states that the force mediated by a light scalar $\phi$ must be stronger than gravitational force and $m^2(\phi)$ is considered as a function of $\phi$ so that $m^2 = V''(\phi)$, then the potential $V(\phi)$ should satisfy
\begin{equation}\label{eq:4.1}
(V^{(3)})^2 \geq \frac{(V^{(2)})^2}{M^2_p}.
\end{equation}  
Furthermore, Eduardo Gonzalo and Luis E.Ibáñez proposed a strong version of SWGC, namely, SSWGC [5]. SSWGC states that the potential of any canonically normalized real scalar $V(\phi)$ in the theory must satisfy for any value of the field the constraint:
\begin{equation}\label{eq:4.2}
2(V^{(3)})^2 - V^{(2)} V^{(4)} \geq \frac{(V^{(2)})^2}{M^2_p},
\end{equation} 
with primes denoting derivation with respect to $\phi$.\\
Higgs inflation model has important phenomenological meaning [6,7]. In our previous work [3], we have found that Higgs inflation model and its two extensions: Higgs-Dilaton model and Palatini Higgs inflation all could satisfy the refining de Sitter swampland conjecture [3].  Furthermore, in this work we intend to explore if Higgs inflation model and its two variations can satisfy the two versions of scalar weak gravity conjecture or not. \\
Based on the Lagrangian of these models, we find that although Higgs inflation model and Higgs-Dilaton model could satisfy the further refining dS swampland conjecture, the two models cannot satisfy SWGC and its strong version. While Palatini Higgs inflation could these conjectures simultaneously. We have obtained the constraints for the Palatini Higgs inflation. \\
Future work can be directed along at least three lines of further research. Firstly, since Palatini Higgs inflation could satisfy two versions of SWGC and the refining de Sitter swampland conjecture simultaneously, it has the potential to be a “real” inflation model of the universe. We should focus on how to derive this inflation model from string theory. Secondly, we should find more inflation models which could satisfy these two kinds of conjectures simultaneously. Thirdly, we should find the common properties of the models which have been found in the second step.






\end{document}